\documentclass[epsfig,graphics,usenatbib]{mn2e}

\usepackage{graphicx}
\usepackage{times}
\usepackage{psfrag}
\usepackage{amsmath,amssymb}
\usepackage{threeparttable}

\title[SFH of ellipticals across the fundamental plane]{The star-formation histories of elliptical galaxies 
across the fundamental plane}
\author[L.A. Nolan et al..]
{L.A. Nolan$^{1}$\thanks{lan@star.sr.bham.ac.uk}, 
J.S. Dunlop$^{2}$, B. Panter$^{2,3}$, Raul Jimenez$^{4,5}$, 
A. Heavens$^{2}$ and G. Smith$^{2}$.
\\
$^{1}$Astrophysics \& Space Science Group, 
School of Physics and Astronomy, University of Birmingham,
Edgbaston, Birmingham, B15 2TT, UK\\
$^{2}$SUPA\thanks{Scottish Universities Physics Alliance}, Institute for Astronomy, University of Edinburgh, Royal Observatory, Edinburgh, EH9 3HJ, UK.\\
$^{3}$Max Planck Institute for Astrophysics, Karl-Schwarzschild-Str. 1, Postfach 1317, D-85741 Garching, Germany\\
$^{4}$Department of Physics \& Astronomy, University of Pennsylvania, 209 South 33rd Street, Philadelphia, PA 19104-6396, USA\\
$^{5}$ The Observatories of the Carnegie Institution, 813 Santa Barbara
St., Pasadena, Ca 91101, USA}

\date{Submitted for publication in MNRAS}

\begin{document}
\newcommand{\Zsolar}{\mbox{$\,\rm Z_{\odot}$}}
\newcommand{\Msolar}{\mbox{$\,\rm M_{\odot}$}}
\newcommand{\Lsolar}{\mbox{$\,\rm L_{\odot}$}}
\newcommand{\xs}{$\chi^{2}$}
\newcommand{\dxs}{$\Delta\chi^{2}$}
\newcommand{\xsn}{$\chi^{2}_{\nu}$}
\newcommand{\ls}{{\tiny \( \stackrel{<}{\sim}\)}}
\newcommand{\gs}{{\tiny \( \stackrel{>}{\sim}\)}}
\newcommand{\asec}{$^{\prime\prime}$}
\newcommand{\amin}{$^{\prime}$}

\maketitle

\label{firstpage}

\begin{abstract}
We present the first results from a study designed to test whether, given high-quality spectrophotometry spanning the mid-ultraviolet--optical 
wavelength regime, it is possible to distinguish the metal content and 
star-formation history of individual elliptical galaxies 
with sufficient accuracy to establish whether 
their formation history is linked to their detailed morphology and position on the 
Fundamental Plane.

From a detailed analysis of ultraviolet-optical spectrophotometry
of the `cuspy' elliptical galaxy NGC 3605 and the giant elliptical
NGC 5018 we find that:
1) optical spectra with $\lambda > 3500$ \AA\ may not contain sufficient information to robustly uncover all the stellar populations present in individual galaxies, even in such relatively passive objects as elliptical galaxies, 
2) the addition of the ultraviolet data
approaching $\lambda = 2500$ \AA\  holds the key to establishing 
well-constrained star-formation histories for these galaxies, from which 
we can infer a formation and evolution history which is consistent with 
their photometric properties, 3) 
despite the superficial similarity of their spectra, the two galaxies have
very different `recent' star-formation histories -- 
the smaller, cuspy elliptical
NGC 3605 contains a high-metallicity population of age $\simeq 1$\,Gyr, 
and has a position on the fundamental plane typical of the product of a 
low-redshift gas-rich merger (most likely at z $\sim$ 0.08), while the giant 
elliptical NGC 5018, with a sub-solar secondary population, appears to have 
gained its more recent stars via mass transfer / accretion of gas from its 
spiral companion, 4) despite these differences in detailed 
history, more than 85$\%$ 
of the stellar mass of both galaxies is associated with an old (9-12\, Gyr) 
stellar population of near-solar metallicity. 

This pilot study provides strong motivation for the construction and analysis 
of high-quality ultraviolet-optical spectra for a substantial sample of 
ellipticals spanning the fundamental plane.

\end{abstract}


\section{Introduction}

Despite recent advances in observational cosmology, the formation and
evolutionary history of elliptical galaxies remains a controversial
subject. Originally thought to be
relatively simple, coeval systems \citep[e.g.][]{f77}, the development of 
hierarchical models in the late 1980s, coupled with the apparent failure
of searches for luminous young ellipticals at high redshift, led many
workers to the viewpoint that most ellipticals have acquired the bulk of
their final mass at relatively low redshift \citep{kc98a}.

In fact, empirical evidence has been presented both for formation of the
bulk of the stars in present-day ellipticals at high redshift \citep{b92,d96,mcc04,c04,r06}, and for the formation of ellipticals from the
merging of spirals at lower redshift \citep[e.g.][]{k96,bcf96,b98,kc98b}. A seemingly anti-hierarchical picture has emerged, with the oldest stellar populations residing in the most massive galaxies \citep[][and references therein]{del06,r06}. 

 Recently, however, this long-running 
debate has taken an interesting turn, with evidence presented that there
exist (at least) two distinct classes of elliptical galaxy, which differ in detailed
morphology and occupy different areas of the fundamental 
plane \citep{kb96,f97}. Specifically, some ellipticals have flattened, or `core' central surface brightness distributions, whereas others have power-law, or `cuspy' central surface brightness distributions. The cuspy ellipticals generally have relatively small effective
radii, intermediate mass, significant rotational support \citep{l95,f97}, and show little or no radio or X-ray emission \citep{b89}. In contrast, the
core ellipticals are generally more massive, have large effective radii,
are pressure supported with very little rotation \citep{l95,f97}, and often show strong radio
emission with high X-ray luminosities \citep{b99}. 

Recent N-body merger simulations indicate that equal-mass mergers, or 
mergers of bulge-dominated systems regardless of the mass ratio, 
lead to slowly rotating systems with flattened central surface brightness 
distributions and `boxy' isophotes, while gas-rich unequal-mass mergers 
produce faster-rotating power-law ellipticals, with `disky' isophotes 
\citep{kb05}. It has been suggested that the core galaxies had the bulk of 
their stars in place a long time ago \citep[e.g.][]{b92}, whereas the 
dissipative mergers producing cuspy ellipticals occurred relatively recently. \cite{g01} have claimed to find direct evidence of the latter process in
action, with their near-infrared study of ULIRGs revealing that these
violent
gas-rich mergers are heading onto the region of the fundamental plane
occupied by power-law ellipticals. This forms an interesting contrast with
the discovery by \cite{d03} that the hosts
of quasars lie in the region of the fundamental plane occupied by giant core 
ellipticals.

Such an apparently anti-hierarchical picture of elliptical galaxy
formation (i.e. in which the most massive objects are the oldest) has
received general support from several recent statistical studies \citep[e.g.][]{k03,crc03,h04}, and 
various theoretical arguments have been advanced to explain how
such a situation could arise within a
CDM universe in which the growth of the underlying dark matter
distribution remains hierarchical. For example, \cite{sr98} argue
that outflows driven by the massive and luminous black hole in a young
massive galaxy could lead to the ejection of gas and termination of star-formation
activity. More recently, \cite{bin03} argues that, in the presence of
sufficiently dense hot atmosphere, cold infalling gas tends to be 
ablated and absorbed by the hot atmosphere, 
preventing any late star-formation activity
beyond that produced by the initial infall of cold gas.
It has also been argued that the flat cores in the most massive
ellipticals may be the result of mergers between two galaxies, each of
which already contains a super-massive black-hole \citep{b80,fm00,hf02}. In this picture, cores could form in the final stages of the merger when the black holes form a binary. Stars are then ejected from the central regions as this binary hardens. Recent N-body simulations support this idea, producing massive core ellipticals from the merging of bulge-dominated galaxies \citep[e.g.][and references therein]{kb05}.

Despite this wealth of observational and theoretical activity, what is
still lacking is a detailed and convincing determination of how elliptical
galaxy star-formation histories vary as a function of position on the
fundamental plane. If this could be properly established, it would be of
enormous assistance in disentangling the relative importance of the
different mechanisms for elliptical galaxy formation currently under
discussion.

That this has not been achieved despite, for example, the availability of
optical spectra for large samples of ellipticals from the Sloan Digital
Sky Survey \citep{e03}, may seem surprising but is due to
two well known problems.
First, there is the problem of age-metallicity degeneracy, with attempts
to determine the ages of stellar populations in elliptical galaxies from
optical data being largely foiled by the fact that colours and absorption
lines (e.g. Mg) are degenerate to compensating changes in age and metal
content \citep[e.g.][]{w94,t99}. Second, although progress has been made 
with the availability of higher-resolution model spectra 
\citep[e.g.][]{bc03}, there is the problem
that with limited spectral range, the data may not allow separation of
different stellar age components. For example, while the Balmer lines are
only moderately sensitive to changes in metallicity, even small amounts of
recent star formation can strengthen the Balmer lines out of all
proportion to the actual mass involved in the burst \citep{t00}.

In this paper we present the results of a pilot study designed to test
whether, given high-quality expanded wavelength coverage extending into the
ultraviolet, these problems can be overcome. Specifically we aimed to test
whether, with accurate spectrophotometry spanning $\lambda \simeq 2500
\rightarrow 8000$ \AA, the age-metallicity degeneracy can be lifted, 
and the
full star-formation history of the elliptical disentangled. 
Various authors have recognised the potential of near-ultraviolet data to
clarify the star-formation history of ellipticals. In particular, \cite{f90} found that mid-UV colour and absorption line indices have
dependences on temperature and metallicity which are strong and distinct
from those derived longward of 3500 \AA\ \citep[see also]{l00}. More
recently, \cite{dcr03} have carried out an extensive
exploration of the usefulness of the mid-UV spectral region for
determining ages and abundances in old stellar populations, and emphasise
the importance of working with low-resolution spectra because of the
present inadequacy of high-resolution models.

We sought, therefore, to construct well-calibrated ultraviolet-optical 
spectra of moderate resolution (i.e. 5-10 \AA) for elliptical galaxies by
combining publically available HST ultraviolet spectroscopy data with
existing optical spectrophotometry. NGC 3605 and NGC 5018 transpire to be 
good representatives of
the cuspy and giant core classes of ellipticals respectively.

The structure of the paper is as follows.
In the next section we briefly 
summarise the known properties of NGC 3605 and NGC 5018, and  
describe the available ultraviolet and optical 
spectroscopic data which can 
be combined to produce 
a properly flux calibrated ultraviolet-optical SED suitable for 
modelling.
Then, in section 3, we describe the model fitting that we have undertaken, 
exploring models of varying sophistication to determine how complex 
the star-formation  histories of these objects could be, and what 
results can be regarded as robust.
In section 4, we present the results of the model-fitting process,
and summarise what can be deduced with confidence 
from this study about the star formation histories of these two 
elliptical galaxies.
Finally, in Section 5, we consider the implications for elliptical 
galaxy formation, and discuss the 
prospects for what could be achieved with a more 
extensive study, before summarising our conclusions in Section 6.

Throughout this paper we assume a flat cosmology with
$H_0 = 70\, {\rm kms^{-1}Mpc^{-1}}$, $\Omega_m = 0.3$, 
and $\Omega_{\Lambda} = 0.7$. The age of the Universe in this model is 13.5 Gyr.

\section{Data}

NGC 3605 is an elliptical galaxy with redshift $z = 0.002228$, and 
absolute magnitude $M_{B} = -18.48$. It is a member of the galaxy group NGC 3607. The UV spectrum ($2260 - 3250$, and $3250 - 4760$ \AA) was observed using the Faint Object Spectrograph (FOS) on the Hubble Space Telescope \citep{petal98}. The optical spectrum ($3520 - 7340$ \AA) was observed with the FAST spectrograph at the F. L. Whipple Observatory's 1.5m Tillinghast telescope \citep{J00}. The FOS observation used a circular 1\asec aperture, and the optical spectrum was observed using a 3\amin x 3\asec slit, scanned over a distance equal to half the blue minor-axis optical diameter.

NGC 5018 is a giant elliptical with redshift $z = 0.009320$ and absolute magnitude $M_{B} = -22.22$ \citep{ss92}. It is a companion to the spiral galaxy NGC 5022. Both an optical \citep{ss92} and an HI \citep{kim88} bridge to this galaxy have been observed. The UV spectrum for this galaxy ($2260 - 4700$ \AA) was also observed using FOS on the Hubble Space Telescope \citep{petal98}. The optical spectrum ($3709 - 8000$ \AA) was observed using the Boller and Chivens spectrograph at the ESO 1.52m telescope at La Silla \citep{BA87}. The final combined spectrum has a total wavelength range  $2260 - 8000$ \AA. Again, the FOS observation used a circular 1\asec aperture. The optical spectrum was observed with a 5\asec x 8\asec slit.

Clearly, the apertures for the UV and optical sections are not matched in either case. We compensate for this in part by allowing the relative normalisation for each spectral segment to float independently where possible (see \S 3). If there has been a significant episode of star formation in the last few gigayears, one would expect the starburst to have occurred in the central regions, so that the smaller FOS aperture may contain flux from a higher proportion of young stars than the larger optical apertures. However, as the UV region of the SED is likely to be dominated by the flux from young stars anyway (thermally-pulsing asymptotic giant branch stars are not significant contributers to the flux at these wavelengths, \citep{mara05}), in this case, the leverage for the ratio of young-to-old stars comes mostly from the optical spectrum, which mitigates the effect of un-matched apertures.

The star-formation rate, estimated from [OII]3727 emission, an error on the FOS fluxes of 6\%, and the relationship derived by \cite{gal99} is $\sim$ (1-a few) $\times$ 10$^{-4}$ \Msolar\ yr$^{-1}$ for both galaxies, which is as one might expect in early-type galaxies such as these.

The errors on the spectra were estimated for each wavelength bin from the 
standard deviation of the nearest nine points to that bin, where the 
bin size was the full width half maximum of the spectral 
observation (8 \AA\ for the UV spectra, 6 \AA\ for the optical 
spectrum of NGC 3605, and 11 \AA\ for the optical spectrum of 
NGC 5018). After the estimation of the errors, the spectra were 
re-binned, where necessary, to either the resolution of the lowest-resolution 
section of the stellar population evolutionary synthesis model spectra 
(10 \AA), or, where the resolution of the data was $>$ 10 \AA, the models 
were binned up to match the data (with errors propagated in quadrature).

\begin{figure}

\includegraphics[width=5.5cm,angle=-90]{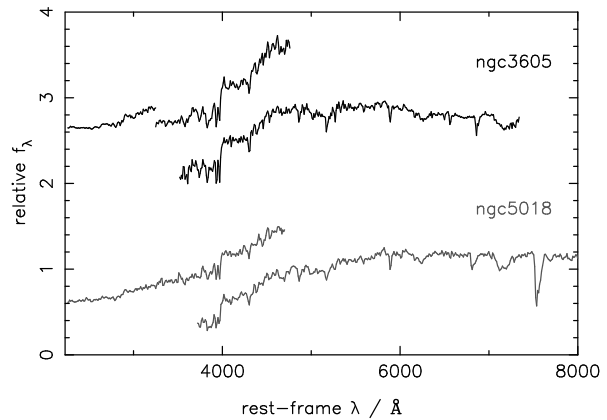}

\caption{\small{The combined UV-optical spectra of the two elliptical 
galaxies NGC 3605 and NGC 5018. For NGC 3605, the spectra have been rebinned to a resolution of 10 \AA. For NGC 5018, the UV portion is at 10 \AA\ resolution, and the optical part at 11 \AA. The normalisation is arbitrarily adjusted for clarity. }}\label{spectra}

\end{figure}

The resulting spectral energy distributions for these two galaxies 
are shown in Figure \ref{spectra}.

\section{Model fitting}	
	
We have undertaken our analysis using the most recently updated models of \citet{bc03}, hereafter BC03. The age range covered by this model set is 0.01 - 14 Gyr\footnote{The resolution of the age grid of the modelled stellar populations is such that the 14 Gyr population is the closest to the age of the universe. For all practical purposes the spectrum at this age can be taken to be equivalent to that of a 13.7 Gyr population.}. The models assume that solar metallicity (\Zsolar) is 0.02, and the available metallicities are: 0.02, 0.2, 0.4, 1.0, 2.5 \Zsolar. The wavelength resolution of our data is higher than the outputs generated by the other modelling codes. Although our data could be rebinned to their resolution, our preferred method is to rebin the models and leave the data its least manipulated state.

The galaxy spectra were fitted with synthetic stellar population models in a 
number of ways. Initially, we have used the MOPED algorithm \citep{h00}, 
which allows the initial assumptions about the star formation history to be 
unconstrained, but does not necessarily return the simplest model consistent 
with the data. We have then fitted each set of galaxy spectra with a single 
stellar population, for each available metallicity in turn, with age and 
relative normalisation allowed to vary freely. Finally, we have fitted 
a two-component model, in which the age, metallicity and relative mass 
fraction of the two components are allowed to vary freely. In the latter two 
cases, the whole parameter space was searched, with the best fit determined 
via \xs\ minimisation. The entire available wavelength range longward of 
2500 \AA\ for each observed galaxy (i.e. 2500 $-$ 7340 \AA\ for NGC 3605, and 2500 $-$ 8000 \AA\ for NGC 5018) was included in the fits.  
The sky lines at $\sim$ 6600 \AA\ and $\sim$ 7560 \AA\ were excluded from the 
fitting in all cases.

\subsection{Fitting with the MOPED algorithm}

In order to determine the optimal multi-population fit for each
spectrum it is necessary to allow as free a star formation history
as possible. A parameterization of 11 single stellar populations, spaced 
logarithmically in
time, each with metallicity varying freely, was allowed, together with a 
one-parameter slab extinction based on the \citet{Gordon} LMC curve. We chose to use the LMC extinction curve as it is currently the most reliably characterised dust screen \citep{Gordon}. Extensive testing in \citet{p06bc} has shown that for non star-burst galaxies the choice of extinction curve does not affect the recovered star formation history at optical wavelengths.

In order to fully assess the resulting 23--dimensional likelihood
surface it was necessary to use the MOPED \footnote{Massively
Optimized Parameter Estimation and Data
compression}\citep{h00, r01} algorithm coupled with a
Markov Chain Monte Carlo search method \citep{Panter03} developed to deal 
with the high resolution BC03 models (Panter et al., in preparation). 
For each galaxy
the best solutions from a series of $100$ randomly-seeded, 
conjugate-gradient searches were used to seed a Markov chain of 
$10^6$ steps.
The individual search traces were examined by eye to confirm full
exploration of the parameter space and estimate solution
convergence. Due to the speed of the MOPED algorithm this analysis
takes of the order three minutes per galaxy on a standard desktop
machine, allowing a thorough investigation of a grid of different
relative normalisations of the UV, visible and IR spectra, in order to 
reproduce the star-formation history as accurately as possible.

For both galaxies the individual sections were scaled to optimally match in the overlap region and the final spectra normalised to unity at 5500\AA\ then analysed with MOPED \citep{p06bc}.

\subsection{Single-metallicity model fits}

Each available, single-metallicity, instantaneous starburst model was fitted 
individually to the spectra of the two galaxies. Each spectral segment 
(three for NGC 3605, and two for NGC 5018, see Figure \ref{spectra}) was 
normalised to a mean flux per unit wavelength of unity across its length. 
For each galaxy, the fit was performed simultaneously on its set of spectra, 
with the model age varying freely. Unlike the MOPED fits, here we are able to allow the normalisation to vary 
independently for each section (to compensate for any deviations in 
the flux calibration). The best-fitting age for each metallicity was 
found via \xs\ minimisation.

\subsection{Two population fits}

Finally, a two-population model was constructed, where the age,
metallicity and fractional contribution by stellar mass for each component were
allowed to vary as free parameters, in order to determine whether more
than one significant episode of star formation has occurred. We have

 \[ F_{2pop,\lambda}  = const( X_{i} f_{Z_{i},\lambda,t_{i}} + X_{j} f_{Z_{j},\lambda,t_{j}} ) \]  

Here, $F_{2pop,\lambda,t}$ is the model flux per unit wavelength in
the bin centred on wavelength $\lambda$ which is the sum of the two
single-metallicity model fluxes (per unit wavelength in the bin
centred on wavelength $\lambda$), $f_{Z_{i},\lambda,t_{i}}$ and
$f_{Z_{j},\lambda,t_{j}}$, which have metallicities and ages, $Z_{i},
t_{i} $ Gyr, and $Z_{j}, t_{j} $ Gyr, respectively. The fractional
contributions of the two components, $X_{i}$ and $X_{j}$, are allowed
to vary freely, and again, normalisation to a mean flux of unity across the wavelength range of each spectral segment was carried out after the
addition of the two unnormalised fluxes. As for the single-component model fits, the two-component fitting was carried out simultaneously for each set of galaxy spectra, with the relative normalisation allowed to vary independently.

A minimum-\xs\  fit was used to determine the best-fit values of
$X_{i}, Z_{i}, t_{i}, X_{j}, Z_{j}$ and $t_{j}$. Again, the whole
parameter space was searched, with the best-fitting parameter values
corresponding to the point on the
age$_{i}$-age$_{j}$-Z$_{i}$-Z$_{j}$ grid with the
minimum calculated \xs.

\section{Results}

\subsection{NGC 3605}

\subsubsection{MOPED result}\label{3605mopedtxt}

\begin{figure}

 \includegraphics[width=5.5cm,angle=-90]{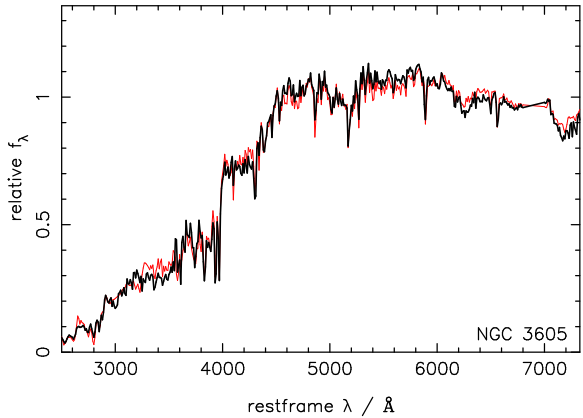}

 \caption{\small Best-fitting model (thin line) to the optimally-spliced spectrum of NGC 3605 (thick line) predicted using MOPED (Heavens et al., 2000). The skylines have been masked out of the fit. The dominant population. has age $=$ 12 Gyr, 1.00 \Zsolar\ and a relative stellar mass fraction of 0.87. There is also a significant (13\% by stellar mass) contribution to the flux from a 1.1 Gyr, 2.32 \Zsolar\ population. See section \ref{3605mopedtxt}
for discussion.}\label{3605mopedfig}

\end{figure}

The spectrum of NGC 3605 was fitted with the MOPED algorithm, and
Figure \ref{3605mopedfig} shows the best­fit result of searching the
23­-dimensional parameter space with optimal relative normalisation of
the UV and optical sections. The results for NGC 3605 are
dominated by two populations with central ages of  1.1 and 12 Gyr and 
metallicity Z $=$ 2.32 \Zsolar\ and Z $=$ 1.00 \Zsolar\ respectively. The
older of the two components accounts for 87\% of the total
stellar population by mass, and the second component accounts for the 
remaining 13\% of the stellar mass. 
The dust attenuation associated with the fit is
$E(B-V)=0.00251$, consistent with the overall picture of an old
elliptical with no recent star formation. The reduced \xs\ of this
fit is 1.774.

\subsubsection{Single-metallicity model fit}\label{36051ztxt}

\begin{figure}
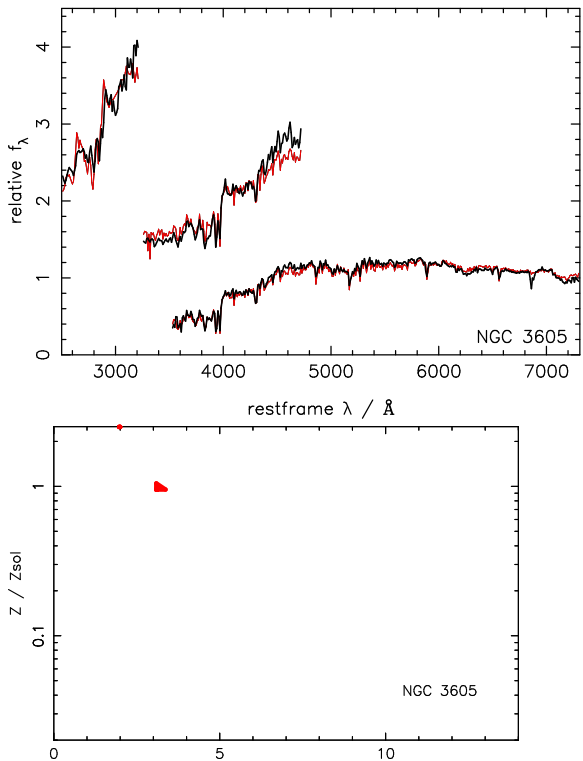


\includegraphics[width=5.5cm,angle=-90]{fig3a.epsf}
\includegraphics[width=4.5cm,angle=-90]{fig3b.epsf}

\caption{\small {\bf Top:} the best-fitting single-metallicity model (1 Gyr, Z $=$ \Zsolar, thin black line) superimposed over the spectra of NGC 3605 (thick black line). {\bf Bottom:} the contours enclose the regions where \dxs $\leq$ 18.4. See section \ref{36051ztxt} for discussion.}\label{36051zfig}

\end{figure}

The results of the single-metallicity fits to the spectra of NGC 3605
are listed in Table \ref{1ztab}. For each metallicity, the free 
parameters are the age and the overall normalisation, which is allowed to 
vary freely for each section to compensate for any inconsistencies in the 
flux calibration of the spectral segments. Figure \ref{36051zfig} shows the 
best-fitting single-population model spectrum, superimposed on the observed 
spectra. The single stellar population model struggles to match the 
shape of the spectrum in the second segment (3300$-$4700 \AA), and 
many of the detailed features. This failure is borne out by an unacceptably 
large value of reduced \xs, namely \xsn = 2.7.

\subsubsection{Two population fit}\label{36052ztxt}

The results for the two-population model fits to the spectrum of NGC
3605 are presented in Table \ref{2ztab} and Figure \ref{36052zfig}. A young, high-metallicity (1 Gyr, 2.5 \Zsolar) component is still found, but the fit is much improved by the addition of a dominant 
mass-fraction of old (10-14 Gyr), lower-metallicity (\Zsolar) stars. Degrading the spectral resolution to 20 \AA\ or 40 \AA\ makes no difference to the recovered best-fit parameters, although \xsn\ decreases to 1.3 and 0.9 respectively.


\begin{figure}

\includegraphics[width=5.5cm,angle=-90]{fig4a.epsf}

\includegraphics[width=7.0cm,angle=-90]{fig4b.epsf}

\caption{\small {\bf Top:} the best-fitting two-component model (thin black line) superimposed over the spectral sections of NGC 3605 (thick black line). 
The two component populations are also shown; the dominant population 
(Z $=$ \Zsolar, age $=$ 14 Gyr, M/M$_{gal} =$ 0.84) is in red, and the 
secondary stellar population (Z $=$ 2.5 \Zsolar, age $=$ 1 Gyr, M/M$_{gal} = $ 0.16) is in green. {\bf Bottom:} the contours enclose the regions where \dxs $\leq$ 16.8. See section \ref{36052ztxt} for discussion.}\label{36052zfig}

\end{figure}

\begin{table}

\begin{center}

\caption{\small Table of results for the single stellar population fits to the spectra of NGC 3605 and NGC 5018. See sections \ref{36051ztxt} and \ref{50181ztxt} for discussion.}\label{1ztab}

\begin{tabular}{lrrr}

\\

\hline

\\

{object} & {Z / \Zsolar} & {age / Gyr} & {min. $\chi^2_{\nu}$} \\

\\
     
\hline

NGC 3605   &  0.02   &  14   &  14.5 \\
           &  0.2    &  12   &   3.5   \\
           &  0.4    &   6   &   3.3  \\
           & {\bf 1.0 }   &  {\bf 3}   & {\bf  2.7}  \\
           & 2.5    &   2   &  2.7  \\

\\

\hline
 
\\

NGC 5018   &  0.02   & 14    &   52.1\\
           &  0.2    & 14    &   11.1\\
           &  0.4    & 14    &    4.2\\
           &{\bf  1.0} &{\bf 10} & {\bf 3.0}\\
           &  2.5    &  3    &    3.2\\

\hline

\end{tabular}

\end{center}

\end{table}

\begin{table}

\begin{center}

\caption{\small Table of results for the two-component stellar population fits to the spectra of NGC 3605 and NGC 5018. See sections \ref{36052ztxt} and \ref{50182ztxt} for discussion.}\label{2ztab}

\begin{tabular}{lrrrr}

\\

\hline

\\

{object} & {Z / \Zsolar} & {age / Gyr} & {M / M$_{gal}$} & {min. $\chi^2_{\nu}$} \\

\hline

NGC 3605                 & 2.5    &   1    &  0.16  &  2.1 \\
                         & 1.0    &  14    &  0.84  & \\
\\

NGC 3605                 & 2.5    &   1    &  0.12  &  2.8 \\
(optical only)            & 1.0    &  5     &  0.88  & \\
\\

\hline
 
\\

NGC 5018                  & 0.4    & 3     &  0.12   & 2.1 \\
                          & 2.5    & 12    &  0.88   & \\
\\

NGC 5018                 & 0.02    &   12    &  0.14  &  2.1 \\
(optical only)            & 2.5   &  14    &  0.86  & \\
\\

\hline

\end{tabular}

\end{center}

\end{table}

\subsection{NGC 5018}

\subsubsection{MOPED result}\label{5018mopedtxt}

The spectrum of NGC 5018 was fitted with the MOPED algorithm, and
Figure \ref{5018mopedfig} shows the best­fitting model resulting from searching the
23­-dimensional parameter space. The results suggest that NGC 5018 is
dominated by a population in the age bin centred on 8.5 Gyr, with metallicity Z $=$ 1.50
\Zsolar\ (94\% by stellar mass), but has a small contribution (5\% by stellar mass)  from the age bin centred on 
2.3 Gyr, with a metallicity of Z $=$  0.79 \Zsolar. There is a very minor (0.07\% by stellar mass) population in the bin centred on 1.1 Gyr, with a solar metallicity. The dust attenuation
associated with the fit is $E(B-V)=0.077$, consistent with the overall picture of an old
elliptical with no recent star formation. The reduced \xs\ of this
fit is 1.39.

\subsubsection{Single-metallicity model fit}\label{50181ztxt}

The results of the single-population fits to the spectrum of NGC 5018
are presented in Table \ref{1ztab}. Figure \ref{50181zfig} shows the observed spectrum, with
the best-fit model spectra superimposed. The minimum reduced \xs\ is \xsn = 
3.0, from which we may infer NGC 5018 cannot be described by a 
single, simple stellar population. The best-fit model has an age 
of 11 Gyr, and solar metallicity. This does not allow the UV flux to be well reproduced.

\subsubsection{Two population fit}\label{50182ztxt}

The results of the two-population fit to the spectrum of NGC 5018 are
presented in Table \ref{2ztab} and Figure \ref{50182zfig}. The addition of a second, younger component (3 Gyr, 0.4 \Zsolar) significantly improves the fit, particularly shortwards of the 4000 \AA\ break. \xsn\ drops from 3.0 for the single stellar population, to 2.1 with the addition of two extra parameters. Degrading the spectral resolution to 20 \AA\ or 40 \AA\ makes no significant difference to the recovered best-fit parameters. \xsn\ decreases to 1.4 and 1.0 respectively, and at 40 \AA\ resolution the fitting recovers a dominant mature population of 11 Gyr, rather than 12 Gyr.


\begin{figure}

\includegraphics[width=5.5cm,angle=-90]{fig5.epsf}

\caption{\small The best-fitting model (thin line) to the optimally-spliced spectrum of NGC 5018 (thick line) predicted using MOPED \citep{h00}. The skylines have been masked out of the fit.  The dominant population. has age $=$ 8.5 Gyr, 1.5 \Zsolar\ and a relative stellar mass fraction of 0.94. There is also a significant (5\% by stellar mass) contribution to the flux from a 2.3 Gyr, 0.79 \Zsolar\ population. See section \ref{5018mopedtxt} for discussion. }\label{5018mopedfig}

\end{figure}


\begin{figure}
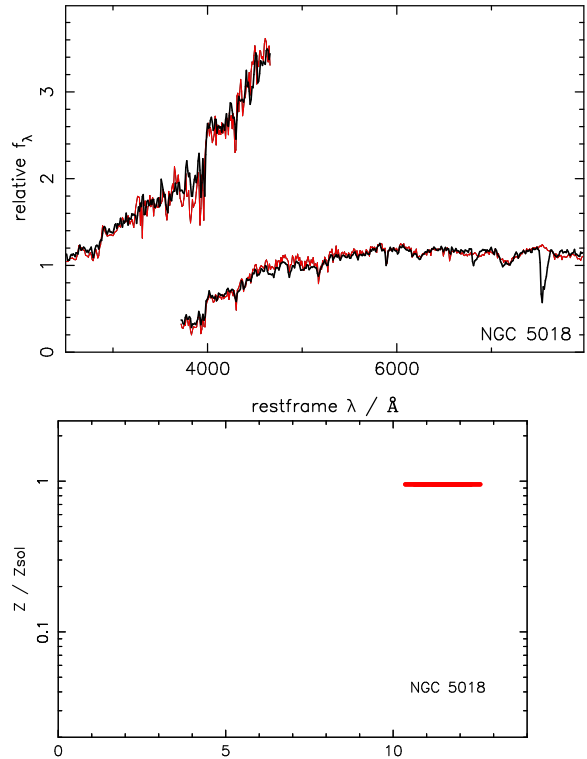


\includegraphics[width=5.5cm,angle=-90]{fig6a.epsf}
\includegraphics[width=4.5cm,angle=-90]{fig6b.epsf}

\caption{\small {\bf Top:} The best-fitting single-metallicity model (thin black line; age $=$ 11 Gyr, Z $=$ \Zsolar) super-imposed over the spectral sections of NGC 5018 (thick black line). {\bf Bottom:} the contours enclose the regions where \dxs $\leq$ 18.4. See section \ref{50181ztxt} for discussion.}\label{50181zfig}

\end{figure}


\begin{figure}

\includegraphics[width=5.5cm,angle=-90]{fig7a.epsf}

\includegraphics[width=7.0cm,angle=-90]{fig7b.epsf}

\caption{\small {\bf Top:} the best-fitting two-component model (thin black line) superimposed over the spectral sections of NGC 5018 (thick black line). The two component populations are also shown; the dominant population (age $=$ 12 Gyr, Z $=$ 0.4 \Zsolar, M/M $_{gal} =$ 0.88) is in red, and the lesser population (age $=$ 3 Gyr, Z $=$ 0.4 \Zsolar, M/M$_{gal} = $ 0.12) is in green. {\bf Bottom:} the contours enclose the regions where \dxs $\leq$ 16.8. See 
section \ref{50182ztxt} for discussion. }\label{50182zfig}

\end{figure}

\section{Discussion}

\subsection{Age and metallicity determination}\label{aztxt}

In this work, model fits to long-baseline spectra have enabled the
age-metallicity degeneracy to be lifted. Comparison of Figures 4 and 8 
(for NGC 3605) and of Figures 7 and 9 (for NGC 5018)
demonstrates the improvement possible in the simultaneous determination of 
both age and metallicity when a wide wavelength range is used. 
In Figures 8 and 9 only the optical (longwards of 3800 \AA) spectra 
have been fitted. Without the leverage from the UV spectra it can be seen 
that the age of the dominant component in NGC 3605 is almost completely 
unconstrained, although the young component is still detected. 
For NGC 5018, although a mature, metal-rich  population is recovered, 
without the extra information provided by the shorter-wavelength data
(which are dominated by younger stars) the fitting process instead tries to compensate in matching the shape of the spectrum by choosing a metal-poor mature population. This clearly demonstrates the power of the long-baseline fit in determining age and metallicity, both in constraining the ages and metallicities of the components, and in selecting the correct components.

\begin{figure}
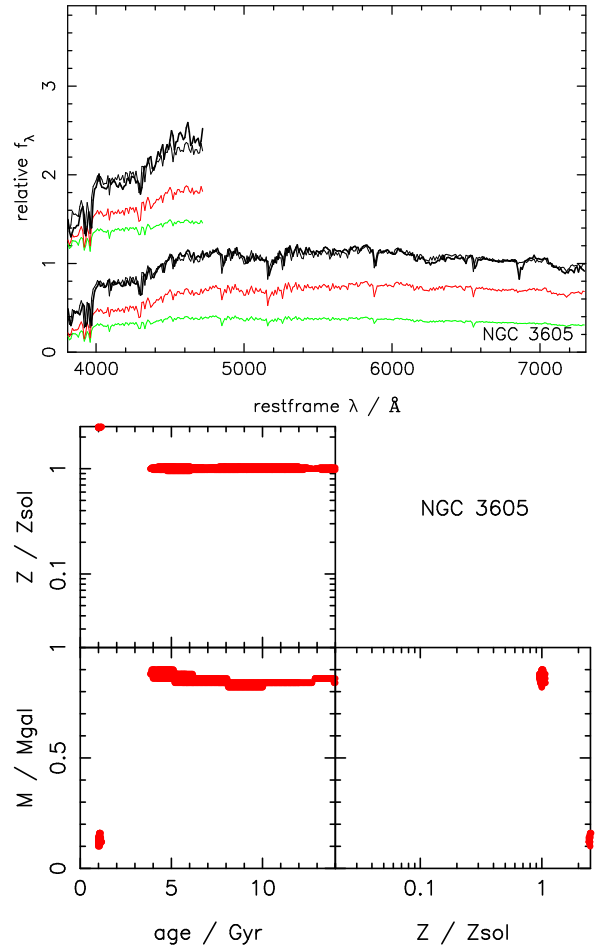


\includegraphics[width=5.5cm,angle=-90]{fig8a.epsf}

\includegraphics[width=7.0cm,angle=-90]{fig8b.epsf}

\caption{\small {\bf Top:} the best two-component model (thin black line) fit to the optical-only spectral sections of NGC 3605 (thick black line). The two component populations are also shown; the dominant population (age $=$ 5 Gyr, Z $=$ \Zsolar, M/M $_{gal} =$ 0.88) is in red, and the lesser population (age $=$ 1 Gyr, Z $=$ 2.5 \Zsolar, M/M$_{gal} = $ 0.12) is in green. {\bf Bottom:} the contours enclose the regions where \dxs $\leq$ 16.8. See 
section \ref{aztxt} for discussion. }\label{3605opfig}

\end{figure}
\begin{figure}

\includegraphics[width=5.5cm,angle=-90]{fig9a.epsf}

\includegraphics[width=7.0cm,angle=-90]{fig9b.epsf}

\caption{\small {\bf Top:} the best two-component model (thin black line) fit to the optical-only spectral sections of NGC 5018 (thick black line). The two component populations are also shown; the dominant population (age $=$ 14 Gyr, Z $=$ 2.5 \Zsolar, M/M $_{gal} =$ 0.86) is in red, and the lesser population (age $=$ 12 Gyr, Z $=$ 0.02 \Zsolar, M/M$_{gal} = $ 0.14) is in green. {\bf Bottom:} the contours enclose the regions where \dxs $\leq$ 16.8. See 
section \ref{aztxt} for discussion. }\label{5018opfig}

\end{figure}

\begin{table}

\begin{center}

\caption{\small Comparison of results for the various stellar population model fits to the spectra of NGC 3605 and NGC 5018. See section \ref{aztxt} for discussion. The agreement between the different models is excellent.}\label{alltab}

\begin{tabular}{lrrrr}

\\

\hline

\\

{model} & {Z / \Zsolar} & {age / Gyr} & {M / M$_{gal}$} & {min. $\chi^2_{\nu}$} \\

\hline

{\bf NGC 3605} & & & & \\
\\

1-component & & &  & \\
BC03                     & 1.0    &   3    &        &  2.7 \\
Maraston (2005)          & 2.0    &   1    &        &  1.9 \\
PEGASE (1997)            & 1.0    &   3    &        &  2.2 \\

\\
 
2-component & & & & \\
BC03                     & 1.0    &  14    &  0.84  & \\
                         & 2.5    &   1    &  0.16  &  \\

Maraston (2005)          & 1.0    & 14     &  0.88  &   1.4\\             
                         & 2.0    &  0.7   &  0.12  &  \\

PEGASE  (1997)           & 1.0    & 14     &  0.94  &  1.4 \\
                         & 1.0    & 1      &  0.06  &  \\

\hline
 
\\
{\bf NGC 5018} & & & & \\
\\

1-component & & & & \\
BC03                     & 1.0    &   10   &        &  3.0 \\
Maraston (2005)          & 0.5    &   14   &        &  3.8 \\
PEGASE (1997)            & 1.0    &   10   &        &  3.4 \\

\\
 
2-component & & & & \\

BC03                      & 2.5    & 12    &  0.88   & 2.1 \\
                          & 0.4    & 3     &  0.12   & \\
Maraston (2005)           & 2.0    & 14    &  0.94   & 2.5   \\
                          & 0.4    & 2     &  0.06   &  \\ 
PEGASE (1997)             & 2.5    & 11    &  0.62   &   2.2\\
                          & 0.2    & 14    &  0.38   & \\

\\

\hline

\end{tabular}

\end{center}

\end{table}

For NGC 3605, all the detailed fitting (BC03: single stellar population, two-component long-baseline and optical-only fitting) agrees that there is a young ($<$ 3 Gyr), metal-rich component  present, but only the long-baseline two-component fitting can constrain the age of the dominant mature population. The results from the full two-component fit are in excellent agreement with the MOPED result.

In Table \ref{alltab}, we compare our one- and two-component fitting using the BC03 models with the best-fitting parameters recovered using models from different authors, namely those of \citet{mara05} and the PEGASE models \citep{Pegase}. These models have a wavelength resolution of 20 \AA\ (lower than the BC03 models) and, although the age grid is the same for all three model sets, the available metallicities are \citet{mara05}: 0.02, 0.5, 1.0, 2.0 \Zsolar, and PEGASE: 0.02, 0.2, 0.4, 1.0, 2.5 \Zsolar. Given the differences between the models, the agreement between the best-fitting parameters recovered for NGC 3605 from the various models is remarkable. The minimum \xs is significantly improved by the addition of a second component, and in all cases, the models find a young ($\leq$ 1 Gyr) secondary component, and a dominant old ($\geq$ 12 Gyr), solar metallicity component. The main difference is that the older PEGASE models do not select the highest available metallicity for the young component, but a solar metallicity instead. However, these results show the robustness of this method to both choice of model and wavelength resolution. The observed 2MASS J-Ks and J-H colours are reproduced by the best-fitting stellar population models to $\leq \pm$ 0.15. This is a reasonable match, as one might expect some discrepancy between the modelled and observed colours, as the observed colours are those for the entire galaxy, whereas the models represent only those stellar populations residing in the central regions of each galaxy.

\citet{tf02} estimated the mean (luminosity-weighted) age of the central region of NGC 3605, by fitting the older stellar population models of \citet{wo97} to H$\beta$ and [MgFe] line indices. They find a mean age of 5.5 Gyr, and a mean metallicity of 1.67 \Zsolar. The high metallicity is consistent with the high metallicity of both our component populations, and the mean age, which may be considered an upper limit to the age of the most recent burst star formation, is also consistent with the presence of the young population that we find.

For NGC 5018, all the detailed BC03 model fits agree that this galaxy is dominated by a mature ($>$ 10 Gyr) metal-rich ($>$ \Zsolar) stellar population. The power of the two-component long-baseline fit allows us, in addition, to disentangle the lesser, younger (3$-$6 Gyr), metal-poor (Z $=$ 0.4 \Zsolar) population. 

Again, we compare our results from the BC03 models to those recovered from the Maraston and PEGASE models in Table \ref{alltab}. Again, the agreement between the various models is remarkable, demonstrating the robustness of this method. Adding a second component significantly improves the fit, and all the models find a dominant old ($\geq$ 12 Gyr) stellar population, at the highest metallicity available. The BC03 and Maraston models agree that there is a sub-solar metallicity, young (2-3 Gyr) component, although again, the older PEGASE models differ, by selecting an old metal-poor population to fit the blue component. The observed 2MASS colours for NGC 5018 are reasonably well-reproduced by all the models, to $\leq \pm$ 0.17.

However, using an independent method (comparing the absorption line indices Ca
II and H$\delta$/$\lambda$4045), and the \citet{w94} models, \citet{lw00} find that a 2.8\,Gyr, \Zsolar\ population 
dominates the spectrum of NGC 5018 at 4000 \AA. They claim an old, metal-poor
population cannot be contributing significantly to the integrated blue
light. Our BC03 and \citet{mara05} fitting results are consistent with this. Although our 3 Gyr, 0.4 \Zsolar\ population only accounts for 0.12 \% of the stellar mass contributing to the spectrum, it dominates the flux at wavelengths $\le 4500$ \AA.

More recently, \citet{bus04}, by matching the HST/FOS UV spectra of NGC 5018 with that of the well-studied compact dwarf elliptical M32, claimed that NGC 5018 is a galaxy whose central regions are dominated by a 3 Gyr stellar population, and that therefore, this galaxy is the child of a major gas-rich merger. They ask the question: `is there any evidence of an older stellar population in NGC 5018, or were literally most of the stars we see formed in the merger event?'. Using our long-baseline fitting technique, we can of course answer that question. We do indeed find a mature stellar population, which dominates the total stellar mass, but, as the young population dominates the flux in the UV, this cannot be detected without using the full spectral range. Conversely, of course, we cannot detect the younger population without including the UV wavelengths.

Although the one- and two-component fitting algorithms do not consider any 
dust extinction of the spectrum, it is clear from the MOPED results
that this is not a large effect in either galaxy. The quality of the
fit over an extended baseline endorses the recovered dust
parameters, as an error in the dust estimation would result in a
gradual offset of the fitting continuum over the length of the
spectrum. The one- and two-component fits suffer less in this respect,
as the relative normalisations of the sections are allowed to vary.

From the detailed fitting, we do not recover formally acceptable values of 
\xs. Recent work \citep[e.g.][]{tmb03} has found super-solar $\alpha$-element abundances present in early-type galaxies. With non-solar abundances calculated for individual features, they are able to recover better and more consistent fits to absorption-line features than for the solar abundances. Non-solar $\alpha$-elements probably account for individual regions of poor fit to the data in this work, but, as long-baseline, non-solar $\alpha$-element stellar population models do not as yet exist, we cannot test whether the quality of the fit will be significantly  improved by their use.

Even though, for both galaxies, the vast majority ($>$ 99\%) of the stars are reproduced by only two stellar populations, the unrestricted star-formation history permitted by the MOPED algorithm, together with its inclusion of a dust parameter, has enabled a better quality of fit to the two low-redshift
elliptical galaxy spectra than that which is possible with even the
best-fitting two-population model, i.e. for NGC 3605, \xsn $=$ 1.8 (compared with 2.1 for the two-component model), and for NGC 5018, \xsn $=$ 1.4 (also 2.1 for the two-component model). 

Despite all of the details discussed above, it is encouraging that, the detailed spectral fits, as well as the 
MOPED fits, are in good agreement regarding the ages of the populations in these galaxies.

\subsection{Mergers and the fundamental plane}

\begin{table}
\begin{center}
\begin{tabular}{lllll}

\hline
{object} & {r$_{e}$ / kpc} & {$\mu_{e}$ (K / mag)} & {$\sigma_{0}$ /
kms$^{-1} $} \\
\hline

NGC 3605 & 1.9  & 17.8 & 103 \\

NGC 5018 & 4.3 & 16.6 & 223 \\

giant core & 7.54 $\pm$ 1.70 & 17.36 $\pm$ 0.68 & 238   $\pm$ 1 \\

power-law & 2.29 $\pm$ 1.75 & 16.75 $\pm$ 1.00 &  157 $\pm$ 2 \\

\hline

\end{tabular}
\caption{\small Observed properties of NGC 3605 and NGC 5018: $\sigma_{0}$ is the central velocity dispersion; $\mu_{e}$, the
effective surface brightness in the K band, is calculated from the
mean surface brightness within the effective half-light radius
(r$_{e}$) in the B band, using ($B - K$) = 3.9 (Genzel et al.,
2001). r$_{e}$ for NGC 5018 is from Scorza et al. (1998), r$_{e}$ for
NGC 3605 is from Faber et al. (1997), and $\mu_{e}$ and $\sigma_{0}$
for both galaxies is from Faber et al. (1989). Also shown are the mean properties of the galaxy samples plotted in 
Figure \ref{fp}. The limits shown are the standard deviation of the sample. All data are taken from Bender et al. (1992), Faber et al. (1997) and Faber et
al. (1989).}\label{obsprop}
\end{center}
\end{table}

\begin{figure}

\includegraphics[width=7.5cm,angle=-90]{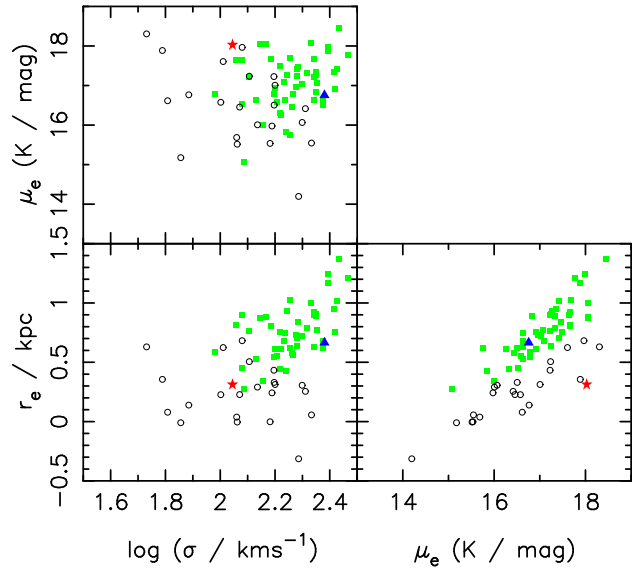}

\caption{\small Projections of the fundamental plane. Green squares: giant / core ellipticals (from Bender et al., 1992, and Faber et al., 1997); black circles: intermediate mass / power-law ellipticals (from Bender et al., 1992, and Faber et al., 1997); red star: NGC 3605 (from Faber et al., 1989 and Faber et al., 1997); blue triangle: NGC 5018 (from Faber et al., 1989 and Scorza et al., 1998). All parameters are calculated for a cosmology with H$_{0} = 70\ {\rm kms^{-1}Mpc^{-1}}$ and $\Omega_{m} = 0.3$. See section 5.3 for discussion. }\label{fp}
\end{figure}

Figure \ref{fp} shows projections in the fundamental plane of the observed
properties of NGC 3605 \citep{f89,f97} and
NGC 5018 \citep{f89,s98}, together with a
population of giant boxy ellipticals, and intermediate mass, disky
ellipticals \citep{b92,f97} for
comparison. The two galaxy populations occupy different positions in
the fundamental plane. Giant ellipticals, with core central surface brightness distributions, tend to be more
massive, more luminous and have higher central velocity dispersions
than the intermediate-mass ellipticals with power-law central surface brightness distributions ('cuspy' galaxies). Mean values of the effective
surface brightness, $\mu_{e}$, effective radius, r$_{e}$, and central
velocity dispersion, $\sigma$, for the two galaxy populations are
listed in Table \ref{obsprop}, together with the corresponding values for NGC 3605 and NGC 5018.

The HST/WFPC2 observations of both \citet{f97} and, more recently,  \citet{l05}, reveal that NGC 3605 is a cuspy galaxy, and its position on the fundamental plane places it unequivocally in the cuspy galaxy population. The spectroscopic determination of the ages of the constituent populations of NGC 3605 is consistent with the theory that cuspy galaxies are the product of a recent gas-rich galaxy-galaxy merger. The bulk of the stars are old ($>$ 10 Gyr), but there is also a significant young (1 Gyr) stellar population, with a higher metallicity. The higher metallicity is expected to result from the burst of star formation triggered in a recent major gas-rich merger, where the new stars are formed from gas which was metal-enriched when the older stars were formed..

MOPED cannot at present fit multiple spectra simultaneously. In this case, therefore, adjoining spectral segments were normalised to unity in the overlap regions, and spliced together.

NGC 5018, a shell galaxy, is at first sight more ambiguous. Its effective radius,
central velocity dispersion and relatively low $\mu_{e}$ suggest that
it belongs with the giant core galaxy population, as one might expect from a brightest group elliptical. At the same time, NGC 5018
appears to contain a fairly substantial young population (12\% by
mass). One might expect this young population to have been triggered
by a merger event. However, NGC 5018 is a near companion to the spiral
galaxy NGC 5022, and there is evidence of an on-going interaction
between these two galaxies: an optical bridge has been detected
between them \citep{ss92}, together with an HI bridge \citep{kim88}. Kim et al. conclude that the HI interaction between
NGC 5018 and NGC 5022 is current, but \citet{mh97} infer from the
deeply embedded dust lane in NGC 5018 that interactions between the
two galaxies have occurred more than once in their
history. Simulations indicate that shell structure resulting from a
weak interaction between galaxies may survive for a few Gyr \citep{red96}. Therefore, although extra blue light from the shells
and bridge mimics a post-merger galaxy, it is more likely that NGC
5018 consists of a predominantly mature central population, which has
evolved passively, together with younger shells which are a product of
mass transfer or accretion. In addition, the lower metallicity of the younger population (0.4 \Zsolar) compared with the older population (2.5 \Zsolar) suggests that the more-recent stars were formed from gas which is either not metal-enriched, or has been diluted by un-enriched gas, which is consistent with star formation triggered by the infall of HI along the bridge between NGC 5018 and NGC 5022. The position of NGC 5018 in the
fundamental plane favours this scenario over that of a major merger between two early-type galaxies.

\section{Conclusions}

We have undertaken a detailed exploration of the star-formation
histories of the elliptical galaxies NGC 3605 and NGC 5018. By assembling and analysing
the spectral energy distributions of these galaxies spanning the wavelength
range 2500 - 8000 \AA\ we aimed to i) determine the added value of
the ultraviolet extension for breaking the age-metallicity
degeneracy, and disentangling star-formation history, and ii) properly constrain and compare the star-formation
histories of these two ellipticals, which occupy very different regions 
of the fundamental plane. Multi-component stellar population fitting allows us to uncover a far more complete picture of the star-formation history of early-type galaxies than fitting to absorption line features.

We find that i) optical spectra with $\lambda > 3500$ \AA\ may not contain sufficient information to robustly uncover all the stellar populations present in individual galaxies, even in such relatively passive objects as elliptical galaxies, 
ii) the addition of the ultraviolet data
approaching $\lambda = 2500$ \AA\  holds the key to uncovering the major epochs of star formation for both galaxies, and establishing 
well-constrained star-formation histories, from which we can infer a formation and evolution history which is consistent with their photometric properties. Our long-baseline fitting technique is remarkably robust to choice of stellar population model and wavelength resolution iii) 
despite the superficial similarity of their spectra, the two galaxies have
very different star-formation histories -- the smaller, cuspy elliptical
NGC 3605 contains a high-metallicity population of age 1 Gyr, and has a 
position on the fundamental plane typical of the product of a low-redshift 
gas-rich merger, while the giant elliptical NGC 5018, with a 
sub-solar secondary population, appears to have gained its more 
recent stars via mass transfer / accretion of gas from its spiral companion, 
iv) despite these differences in detailed 
history, more than 85$\%$ 
of the stellar mass of both galaxies is associated with an old (9-12\, Gyr) 
stellar population of near-solar metallicity.

For a universe with $\Omega_{m}$ = 0.3, $\Omega_{\Lambda}$ = 0.7 and
H$_{0}$ = 70 kms$^{-1}$Mpc$^{-1}$, it thus appears 
that the redshift at which NGC 3605 was
assembled to its present state is z$_{merg} =$ 0.076, if the
formation of the stars in the 1 Gyr population is assumed to have
been triggered by this event. This is in agreement with hierarchical
structure formation scenarios in which disk-disk merging events take place
predominantly at z $<$ 1. However the 
vast majority of its stars appear to have formed 
at $z _f > 3$. NGC 5018 also formed the vast 
majority of its stars 
at high redshift (z$_{f} = 2 - 5$), but then appears to have gained a shell 
(or shells) of 
younger stars between z $ \simeq 0.3$ and the present-day, possibly 
through interaction with NGC 5022.

In conclusion, the spectroscopic disentanglement of the component
populations of elliptical galaxies is possible, and is a useful tool
in the investigation of elliptical galaxy formation and evolution. We
are able to determine the formation scenarios of these galaxies, and
hence their general position in the fundamental plane using this
method. The availability of full SEDs with non-solar $\alpha$-element abundances would help to further improve the goodness-of-fit achieved with this method.

\section*{ACKNOWLEDGEMENTS}
This work was based in part on 
observations with the NASA/ESA {\it Hubble Space Telescope},
obtained at the Space Telescope Science Institute, which is operated by the 
Association of Universities for Research in Astronomy, Inc. under NASA 
contract No. NAS5-26555.
Louisa Nolan acknowledges the support of PPARC, through the award of a PDRA.
Raul Jimenez is supported by NSF grants AST-0408698 and PIRE-0507768, and NASA grant NNG05GG01G. Ben Panter is supported by the Alexander von Humboldt
Foundation, the Federal Ministry of Education and Research, and the
Programme for Investment in the Future (ZIP) of the German Government.

\end{document}